\documentclass[]{aastex631}
\usepackage{array}
\usepackage{tabularx}
\usepackage{amsmath} 

\submitjournal{ApJ}

\shorttitle{Frequency Dispersed Ion Acoustic Waves}
\shortauthors{Malaspina et al.}

\begin{document}

\title{Frequency Dispersed Ion Acoustic Waves in the Near Sun Solar Wind: Signatures of Impulsive Ion Beams}

\correspondingauthor{David Malaspina}
\email{David.Malaspina@colorado.edu}

\author[0000-0003-1191-1558]{David M. Malaspina}
\affiliation{Astrophysical and Planetary Sciences Department, University of Colorado, Boulder, CO, USA}
\affiliation{Laboratory for Atmospheric and Space Physics, University of Colorado, Boulder, CO, USA}

\author[0000-0002-3096-8579]{Robert E. Ergun}
\affiliation{Astrophysical and Planetary Sciences Department, University of Colorado, Boulder, CO, USA}
\affiliation{Laboratory for Atmospheric and Space Physics, University of Colorado, Boulder, CO, USA}

\author[0000-0001-6978-9765]{Iver H. Cairns}
\affiliation{School of Physics, University of Sydney, Sydney NSW 2006, Australia}

\author[0000-0003-3945-6577]{Benjamin Short}
\affiliation{Department of Physics, University of Colorado, Boulder, CO, USA}
\affiliation{Laboratory for Atmospheric and Space Physics, University of Colorado, Boulder, CO, USA}

\author[0000-0003-1138-652X]{Jaye L. Verniero}
\affiliation{NASA Goddard Space Flight Center, Greenbelt, MD, USA}

\author[0000-0002-3805-320X]{Cynthia Cattell }
\affiliation{School of Physics and Astronomy, University of Minnesota, Minneapolis, MN 55455, USA}

\author[0000-0002-0396-0547]{Roberto Livi}
\affiliation{Space Sciences Laboratory, University of California, Berkeley, CA, USA}

\begin{abstract}

This work reports a novel plasma wave observation in the near-Sun solar wind: frequency-dispersed ion acoustic waves.  Similar waves were previously reported in association with interplanetary shocks or planetary bow shocks, but the waves reported here occur throughout the solar wind sunward of $\sim 60$ solar radii, far from any identified shocks.  The waves reported here vary their central frequency by factors of 3 to 10 over tens of milliseconds, with frequencies that chirp up or down in time.  Using a semi-automated identification algorithm, thousands of wave instances are recorded during each near-Sun orbit of the Parker Solar Probe spacecraft.  Wave statistical properties are determined and used to estimate their plasma frame frequency and the energies of protons most likely to be resonant with these waves.  Proton velocity distribution functions are explored for one wave interval, and proton enhancements that may be consistent with proton beams are observed.  A conclusion from this analysis is that properties of the observed frequency-dispersed ion acoustic waves are consistent with driving by cold, impulsively accelerated proton beams near the ambient proton thermal speed.  Based on the large number of observed waves and their properties, it is likely that the impulsive proton beam acceleration mechanism generating these waves is active throughout the inner heliosphere. This may have implications for acceleration of the solar wind.  

\end{abstract}



\section{Introduction} 
\label{sec:intro}

Ion instabilities that drive plasma waves exist in many regions of the solar wind.  At magnetohydrodynamic (MHD) scales, ion cyclotron and firehose instabilities constrain the allowed parameter space for temperature evolution of the solar wind \citep{Kasper2002, Hellinger2006, Matteini2007}.  Coherent ion cyclotron wave packets appear throughout the solar wind far from interplanetary shocks or planetary bow shocks, sometimes appearing in storms \citep{Jian2009, Jian2014, Boardsen2015}.  They are particularly prevalent in the near-Sun solar wind \citep{Bowen2020, Verniero2020, Carbone2021}, driven by perpendicular temperature anisotropy (\citet{Verscharen2019} and references therein), or ion beams \citep{Wicks2016, Verniero2022}.  Ion distributions with beam features or shoulders in the direction of the background magnetic field are a persistent feature of the near-Sun solar wind \citep{Marsch1982, McManus2024}. 

Higher frequency waves, including ion acoustic waves, are well-known to occur near magnetic discontinuities such as interplanetary shocks \citep{Balikhin2005, Cohen2020, Davis2021} and near planetary foreshocks \citep{Anderson1981, Fuselier1984, Hull2006, Wilson2014, Goodrich2019, Vech2021, Vasko2022}, where reflected ion beams create unstable distribution functions \citep{Filbert1979, Eastwood2005}.  Plasma waves in the solar wind with unusual properties (repetitive wave packet modulation, very long durations, quasi-monochromatic) have been reported in the near-Sun solar wind, far from any shock-like structure, and were identified as ion acoustic waves \citep{Mozer2021} with nonlinear growth characteristics \citep{Mozer2023}.  

This work reports on a newly-identified type of ion acoustic wave in the near-Sun solar wind (sunward of $60$ solar radii, $R_S$): the frequency-dispersed ion acoustic wave. These waves are polarized along the background magnetic field direction and have no measurable magnetic component.  They rapidly change their central frequency, chirping from low to high, or high to low, frequency by factors of 3 to 10 over small fractions of a second.  Their plasma frame frequency is somewhat below the local ion plasma frequency.  They often appear in clusters which may persist from a few seconds to hours.   Ion-acoustic like waves with similar frequency-variability were identified by \citet{Cohen2020} and \citet{Davis2021}, but in those cases the waves were downstream of an interplanetary shock.  By contrast, the waves reported here are observed far from interplanetary shock fronts or other magnetic discontinuities known to generate proton beams.  Finally, the identified waves are abundant in the near-Sun solar wind, appearing on every orbit of the Parker Solar Probe spacecraft, with up to tens of thousands of wave detections per orbit. 

This work examines properties of these frequency-dispersed ion acoustic waves in the near Sun solar wind, including their plasma-frame frequency and the velocity of wave-resonant protons.  Proton distribution functions are examined and found to show features consistent with proton beams when the frequency-dispersed ion acoustic waves are present. Wave properties are found to be consistent with beam-mode ion acoustic waves, likely driven by cold impulsive proton beams.  

While the mechanism generating the proton beams is beyond the scope of this work, the prevalence of the observed frequency-dispersed ion acoustic waves suggests that the proton beam generation mechanism must occur through a large portion of the near-Sun solar wind.  Such pervasive proton acceleration may have implications for global solar wind acceleration, particularly if the global ambipolar electric field identified by \citet{Bercic2021} is composed of Debye-scale electric potential structures.

\section{Data} 
\label{sec:data}

This study uses data from the Parker Solar Probe spacecraft \citep{Fox2016} during 15 of its first 16 close encounters with the Sun.  Data from the FIELDS \citep{Bale2016} and SWEAP \citep{Kasper2016} instrument suites are used.  Data from radial distances sunward of $\sim70$ solar radii ($R_{S}$) are considered in this study. Data from Encounter 1 are not examined due to the FIELDS instrument channel configuration during that encounter.

FIELDS data in this study include AC-coupled electric field data from the four antennas in the plane of the spacecraft heat shield, AC-coupled magnetic field data from the search coil magnetometer (SCM), and DC-coupled magnetic field data from the fluxgate magnetometer.  The three-axis fluxgate data are continuously sampled at a cadence $> 73$ samples/s.  The electric field and SCM data analyzed here include on-board calculated power spectra from the Digital Fields Board (DFB) \citep{Malaspina2016} covering $\sim365$ Hz to to 75 kHz.  Electric field and SCM time series data recorded by the DFB burst capture algorithm \citep{Malaspina2016} are also examined. These data are sampled at $150,000 / 2^N$ samples per second, where N is an integer between 0 and 3.  N may change between encounters or within an encounter, but it may not change during any recorded burst interval.     

SWEAP data used include proton distribution functions from the SPAN-Ion instrument (SPANi) \citep{Livi2022}, as well as proton distribution function moments (velocity, density, temperature) from both SPANi and the Solar Probe Cup (SPC) Faraday cup \citep{Case2020}. 

\section{Ion Acoustic Wave Observations} 
\label{sec:waves}

\begin{figure}[ht!]
\includegraphics[width = 89mm]{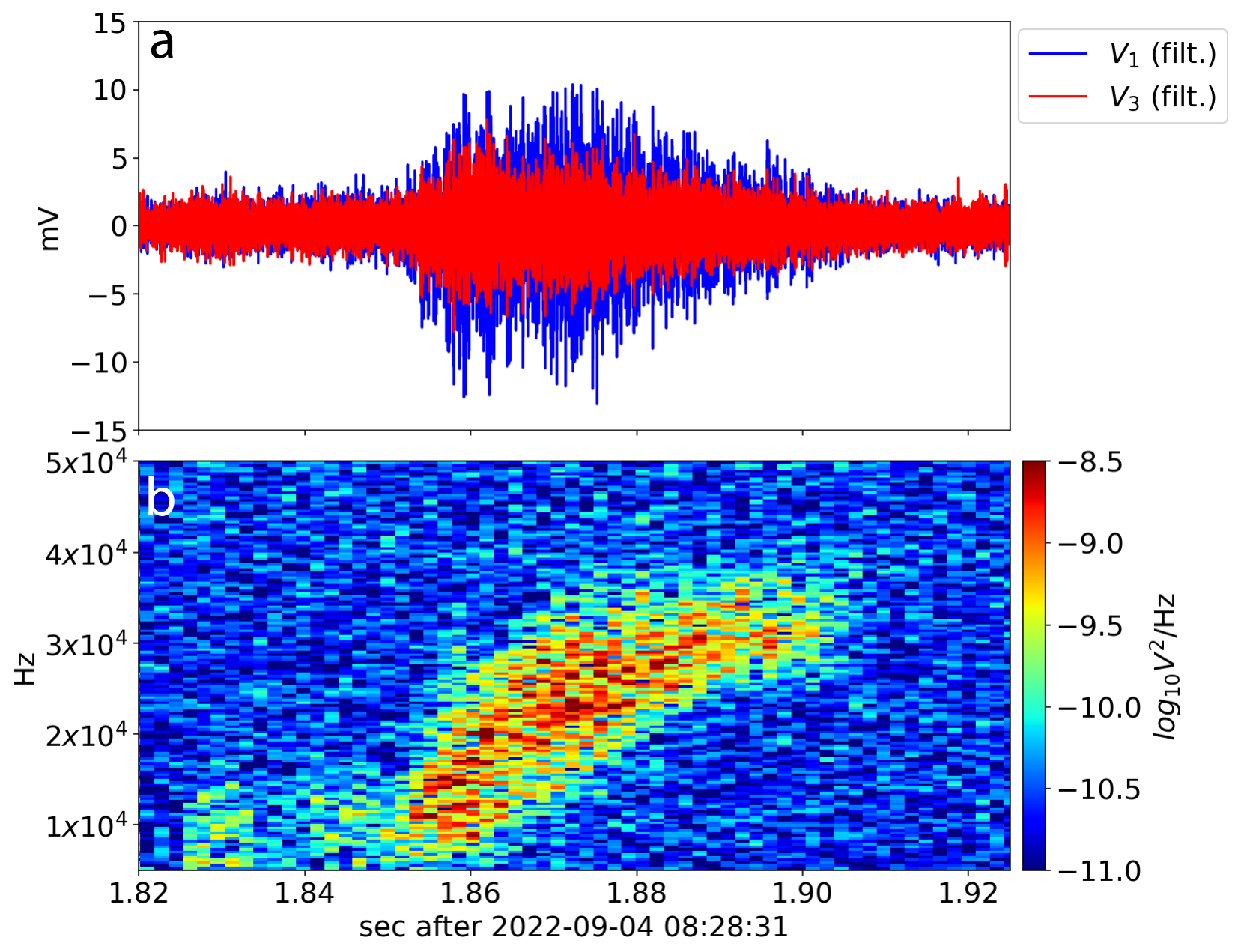}
\caption{Example of an individual time-dispersed ion acoustic wave packet, observed using high resolution burst data.  (a) Time series of single-ended antenna voltage signals from two nearly orthogonal antennas, bandpass filtered as described in the text, (b) Windowed Fourier transform of the non-filtered time series data from this interval. }
\label{fig_01}
\end{figure}

Figure \ref{fig_01} shows a plot of an isolated time-dispersed ion acoustic wave packet, recorded on 4 September 2022 near 08:28:31 UTC using high resolution DFB burst data.  Parker Solar Probe was $26.9 R_S$ from the Sun at this time.  Single-ended voltage differences between electric field antennas and the spacecraft chassis ground $V_{1} = V_{Ant \ 1} - V_{sc}$ and $V_{2} = V_{Ant. \ 2} - V_{sc}$ are plotted. To enable visual identification of wave packets, the time-series data in Figure \ref{fig_01}a are bandpass filtered using 5th order Bessel filters, with 3 dB points at 5 kHz and 50 kHz.  Figure \ref{fig_01}b shows a windowed Fourier transform of the non-filtered time series data from this interval. A linear frequency axis is used to illustrate the rapid upward chirping frequency variation in the wave power.  There is no measurable wave power or spectral signature in the SCM data at this time.  

\begin{figure}[ht!]
\includegraphics[width = 89mm]{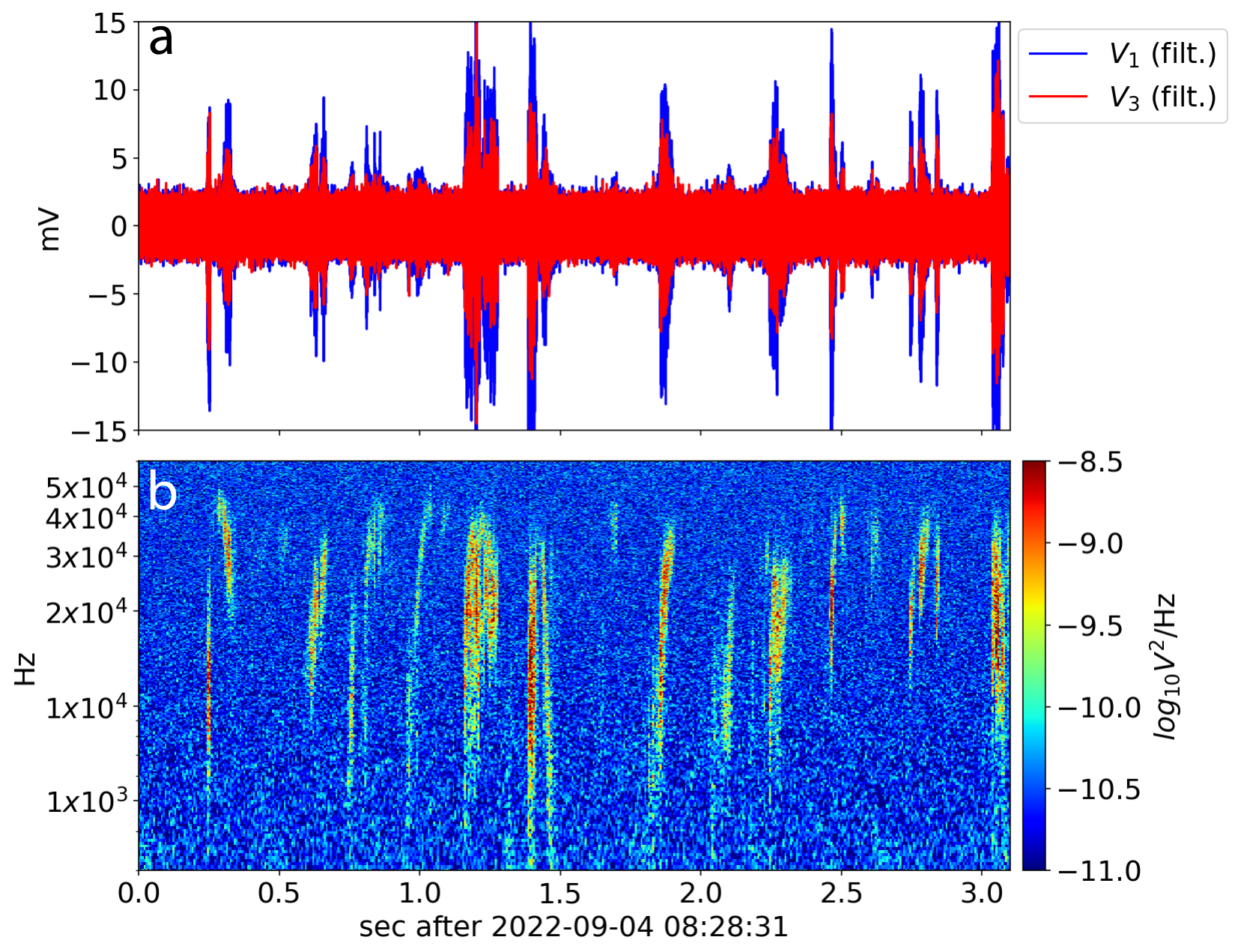}
\caption{ A few-second interval with many time-dispersed ion acoustic wave packets, using the same format as Figure \ref{fig_02}, but with a broader frequency axis in (b) compared to Figure \ref{fig_01}b. }
\label{fig_02}
\end{figure}

The wave power changes frequency at a rate of 625,000 Hz / s.  This rate of frequency change is most likely too rapid to be caused by changing solar wind conditions alone. The solar wind proton velocity, temperature, and density are determined by SPANi to be $305.7$ km/s (in the spacecraft frame), $35.4$ eV, and 612 $cm^{-3}$.  Therefore, the 0.04 s duration of the wave packet corresponds to 12.2 km of plasma passing the spacecraft.  The calculated ion inertial length at this time is $\sim9.2$ km.  If the frequency change were entirely due to change in the solar wind speed, the solar wind speed would need to increase by a factor of $\sim6$ over 1 ion inertial length and just as abruptly return to $\sim300$ km/s (see Figure \ref{fig_02}).  A more likely explanation is that the shift in frequency is intrinsic to the wave growth and/or propagation characteristics. 


While the wave packet in Figure \ref{fig_01} is isolated and well-defined, these wave packets often appear in clusters, where individual wave packets may overlap.  Figure \ref{fig_02} shows the full $\sim3$ s burst interval that contains the wave packet shown in Figure \ref{fig_01}.  The time series data in Figure \ref{fig_02}a are filtered the same way as in Figure \ref{fig_01}a. In the full burst capture, many wave packets are visible, most with rising tone frequency chirps.  The packet near 0.3 s shows a falling tone, as do several others.   Again, there is no corresponding wave power or spectral signature in the SCM data.  


The rapidly varying frequency structure of these frequency-varying ion acoustic waves is observable in high-resolution time series burst captures ($> 18,750$ sample/s).  These same waves also appear in the higher frequency on-board processed DFB survey spectra (dfb\_ac\_spec data product).  However, the survey spectra cannot resolve the rapidly varying frequency structure of each wave packet due to the time-averaging, frequency-averaging, and duty-cycling applied to the on-board processed data.  The higher frequency DFB survey spectra cover frequencies from $\sim360 $ Hz to $75$ kHz.  In on-board processing, the electric field antenna and SCM data are time-averaged over $\sim$0.11 s and frequency-averaged into 56 pseudo-logarithmically spaced bins. For each channel sampled, a spectra containing $\sim$0.11 s of data, is reported every $\sim$0.87 s. 

\begin{figure}[ht!]
\includegraphics[width = 140mm]{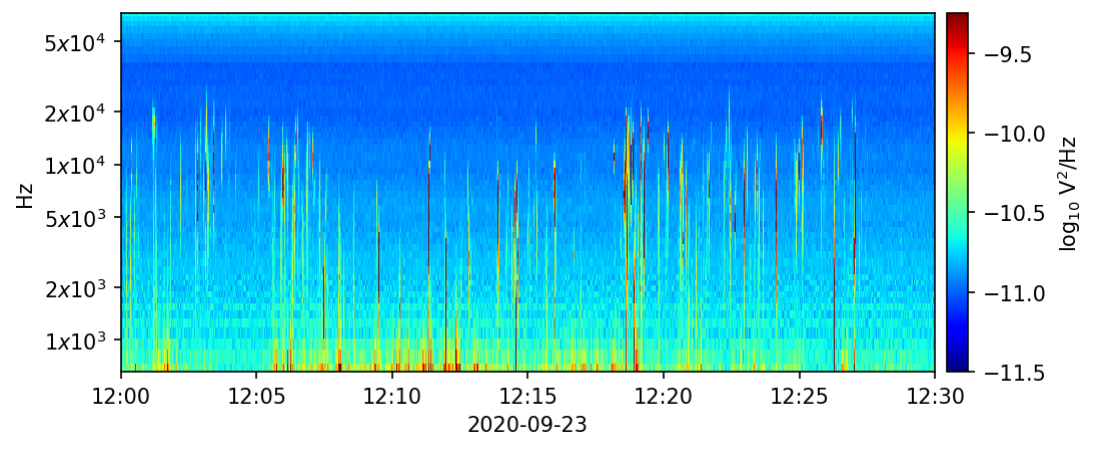}
\caption{ A long interval ($\sim 30$) minutes of with many time-dispersed ion acoustic wave packets, observed using survey spectral data from the two differential electric field antenna pairs.}
\label{fig_03}
\end{figure}

As a result of this processing, the frequency-varying ion acoustic waves appear in survey spectra as brief, broadband impulses.  Figure \ref{fig_03} shows an example of a spectrogram containing these waves.  The summed spectra from two differential electric field channels $V_{12} = V_{Ant. \ 1} - V_{Ant. \ 2}$ and $V_{34} = V_{Ant. \ 3} - V_{Ant. \ 4}$ are plotted. Again, no corresponding signal is observed in the SCM spectra.  The waves appear in sporadic patches for the entire 30 minute interval shown.  While analysis of the high resolution burst data show that any given wave packet has a variable bandwidth and amplitude (e.g. Figure \ref{fig_01} and Figure \ref{fig_02}), the spread in these quantities is exaggerated by the on-board processing.  `Storms' of frequency-varying ion acoustic waves (e.g. Figure \ref{fig_03}) are commonly observed by Parker Solar Probe, as are isolated instances of these waves.


\subsection{Wave Identification} 
\label{sec:event_identification}

To investigate the behavior of these waves over many encounters and across solar wind plasma conditions, we developed an algorithm to automate their identification in survey spectral data.  

First, all intervals with $V_{12}$ and $V_{34}$ AC-coupled spectral survey data at $0.87$ s cadence are identified. These data are recorded sunward of 60 $R_S$ for most solar encounters, and they are recorded at higher altitudes for some encounters.  $V_{34}$ AC-coupled spectral survey data were not recorded on encounter 1. Therefore encounter 1 is excluded from this analysis.  The $V_{12}$ and $V_{34}$ data are summed at each time step to make a combined differential voltage power spectra ($V_{T}$). The combined data are then broken up into 15 minute intervals, and the following algorithm is applied to each interval.  

For each interval, $V_{T}$ is divided by the maximum power spectral value in that interval, then multiplied by 256.  Sobel transforms are then applied in the vertical (frequency) and horizontal (time) directions separately.  Sobel transforms convolve a kernel with an image in order to highlight gradients in that image. The vertical ($K_{Sv}$) and horizontal ($K_{Sh}$) kernels used are: 

\[
K_{Sv} =
\begin{bmatrix}
 1 & 2 & 1\\
 0 & 0 & 0\\
-1 & -2 & -1
\end{bmatrix}
\ \ \
K_{Sh} =
\begin{bmatrix}
 1 & 0 & -1\\
 2 & 0 & -2\\
1 & 0 & -1
\end{bmatrix}
\]

After the Sobel transforms are applied, the resulting vertical and horizontal gradient-detection images are $I_{v}$ and $I_{h}$.  Next, vertical and horizontal image thresholds are determined ($T_{v}, T_{h}$), defined as 5 times the upper quartile of the absolute value of all pixels in $I_{v}$ and $I_{h}$, respectively.  A binary vertical image mask is created by retaining pixels ($mask_v = 1$) where $|I_{v}| > T_{v}$.  A binary horizontal image mask is created by retaining pixels ($mask_h = 1$) where $I_{v}(t-1,f) < -T_{v}$ and $I_{v}(t+1,f) > T_{v}$, for a given time $t$ and frequency $f$. This vertical image mask selects for strong vertical gradients, and the horizontal image mask selects strong gradients that are isolated in time at each frequency.  The vertical and horizontal masks are then combined with a bitwise AND operation, requiring both conditions to be true for a pixel to be retained ($mask = mask_h \ AND \ mask_v$).  

After this initial wave identification mask is created, additional processing is required to remove lower frequency ($< 1 $ kHz) broadband waves \citep{Malaspina2022_KAW} and dust impact signatures \citep{Malaspina2022_dust} from the data.  Two additional criteria are applied:  (i) For each time (t), the lowest frequency bin where $I_{v}(f,t) < T_{v}$ is identified ($f (i_{low1})$) for frequency bin index $i_{low1}$).  All data at that time in frequency bins near or below this frequency are removed from consideration ($mask(f,t) = 0$ for $f < f (i_{low1}+4) $).  (ii) For each time (t), the lowest frequency bin where the amplitude drops below a threshold (5 times the median amplitude across time in the 67,968.75 Hz frequency bin) is identified ($f (i_{low2})$), for frequency bin index $i_{low2}$.  All data at this time at frequencies near or below $f (i_{low2})$ are removed from consideration ($mask(f,t) = 0$ for $f < f (i_{low2}+4) $).  This process eliminates signals that are strongest in the lowest frequency bins and drop off toward higher amplitudes, which is a characteristic behavior of low frequency broadband waves and dust impact signatures.   Eliminating data from the four frequency bins above $f(i_{low1})$ and $f(i_{low2})$ is important because the amplitude gradient is often ragged and non-uniform at the high frequency edge of the low frequency waves.  

Finally, the resulting image mask is multiplied by the original differential power spectra, setting most pixels to 0.  At each time, the frequency bin of the masked spectra with the largest non-zero amplitude is identified as the likely wave.  The wave power selected this way corresponds to inflections between positive and negative vertical gradients in the power spectra.  The time, frequency, and amplitude of each wave identified through this automated procedure are then recorded.   

\begin{figure}[ht!]
\includegraphics[width = 95mm]{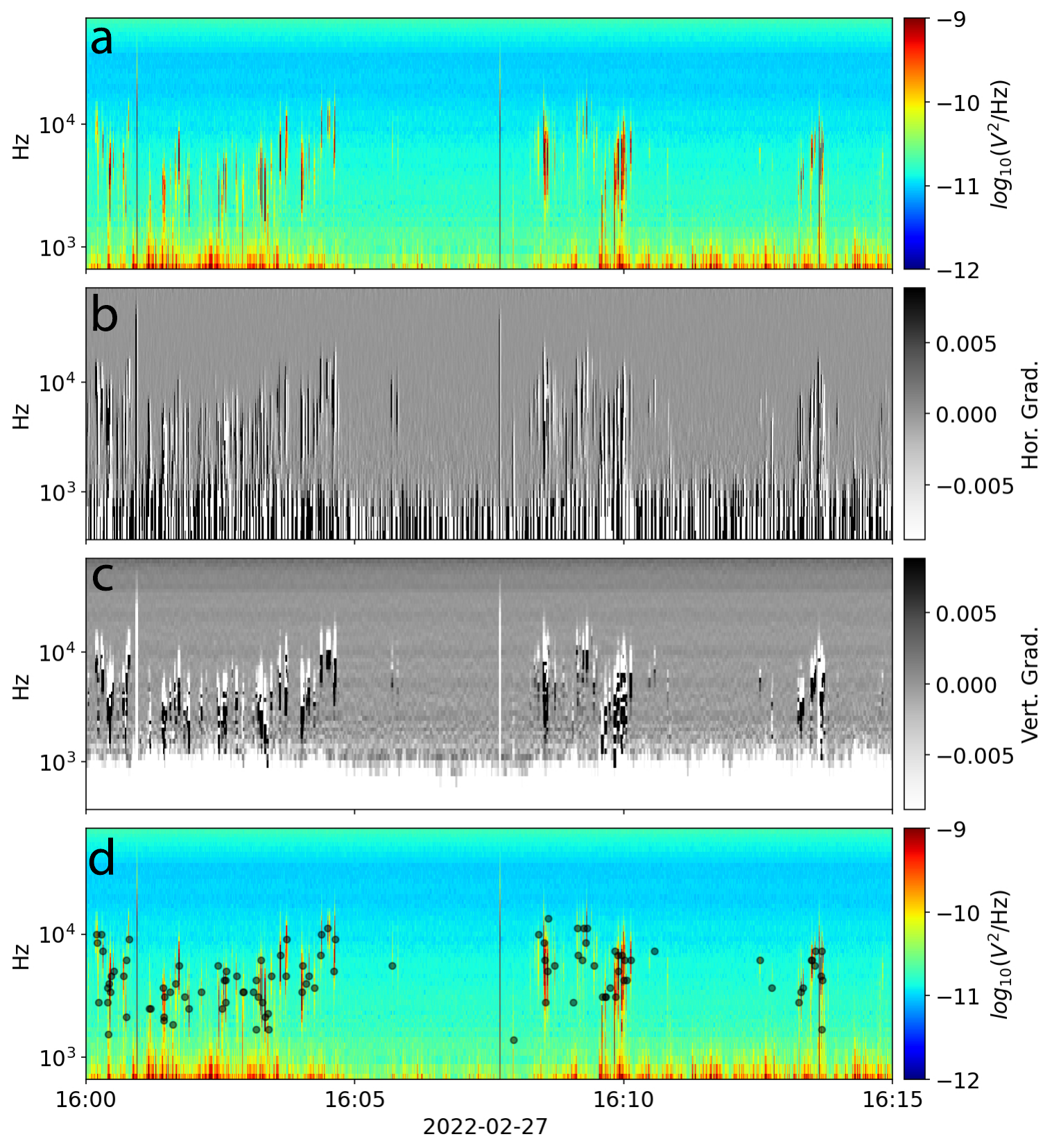}
\caption{ Illustration of the method used to automatically identify frequency-dispersed ion acoustic waves in survey spectral data.  (a) Survey spectral data for a 15 minute interval from the two differential electric field antenna pairs.  (b) Data from (a), after amplitude scaling and application of the horizontal Sobel transform (see text for detail).  (c) Data from (a), after amplitude scaling and application of the vertical Sobel transform.  (d) Data from (d), with the frequency and time of the peak amplitude for each identified frequency-dispersed ion acoustic wave indicated by a black dot.}
\label{fig_04}
\end{figure}

Figure \ref{fig_04} shows the progression of these processing steps.  Figure \ref{fig_04}a shows the combined differential voltage power spectra ($V_{T}$) for a 15 minute interval on Feb. 27, 2022, during encounter 11. Parker Solar Probe was $\sim28 R_S$ from the Sun at this time.  Spectral signatures of the time-dispersed ion acoustic waves are visible.  Both low frequency broadband waves ($< 1 kHz$) and dust impact signatures (near 16:01 and 16:07 UTC) are also visible.  Figure \ref{fig_04}b and Figure \ref{fig_04}c show $I_{h}$ and $I_{v}$, the gray scale images after the horizontal and vertical (respectively) Sobel filters are applied.  The color scale limits are set to $\pm T_{h}$ and $\pm T_{v}$ for Figure \ref{fig_04}b and Figure \ref{fig_04}c, respectively.  Figure \ref{fig_04}d repeats the power spectral plot from Figure \ref{fig_04}a, but with a black dot indicating each identified wave.  Low frequency broadband waves and dust impacts are successfully removed from consideration during this interval.  


This automated wave-detection algorithm is then applied to all Parker Solar Probe data from encounters 2 through 16.  If a 15-minute interval has 10 or fewer wave detections, it is removed from consideration.  The resulting wave identifications are also examined by eye to remove wake-generated waves near the local electron cyclotron frequency \citep{Tigik2022, Malaspina2022_wakes}.  The wake-generated waves are readily identified as near-harmonic bands of power at approximately constant frequency close to the electron cyclotron frequency.   While it may be possible to remove the wake-generated waves algorithmically, it was more time-efficient for this study to remove them by visual inspection.  

For each identified wave event, solar wind plasma conditions are recorded, including the magnetic field vector and the proton distribution function moments: plasma density, proton temperature, and solar wind velocity.  Proton moments are taken from SPANi data whenever possible, and from SPC data otherwise. SPANi data are used when the peak of the proton distribution function occurs in any detector anode except the two closest to the heat shield.  This criteria minimizes errors in SPANi proton distribution function moment estimates due to due to partial field of view coverage \citep{Verniero2020}.  The trace of the proton temperature tensor is used to represent proton temperature. For both SPANi and SPC data, the median value of each proton moment over the 15 minute interval is assigned to all waves identified in that interval. 

\subsection{Orbital Trends} 
\label{sec:orbital_trends}

A database of frequency-dispersed ion acoustic waves was created by applying wave identification algorithm described in the prior section to data from Parker Solar Probe's encounters 2 through 16.  Here, that database is examined for orbital trends.  

\begin{figure}[ht!]
\includegraphics[width = 89mm]{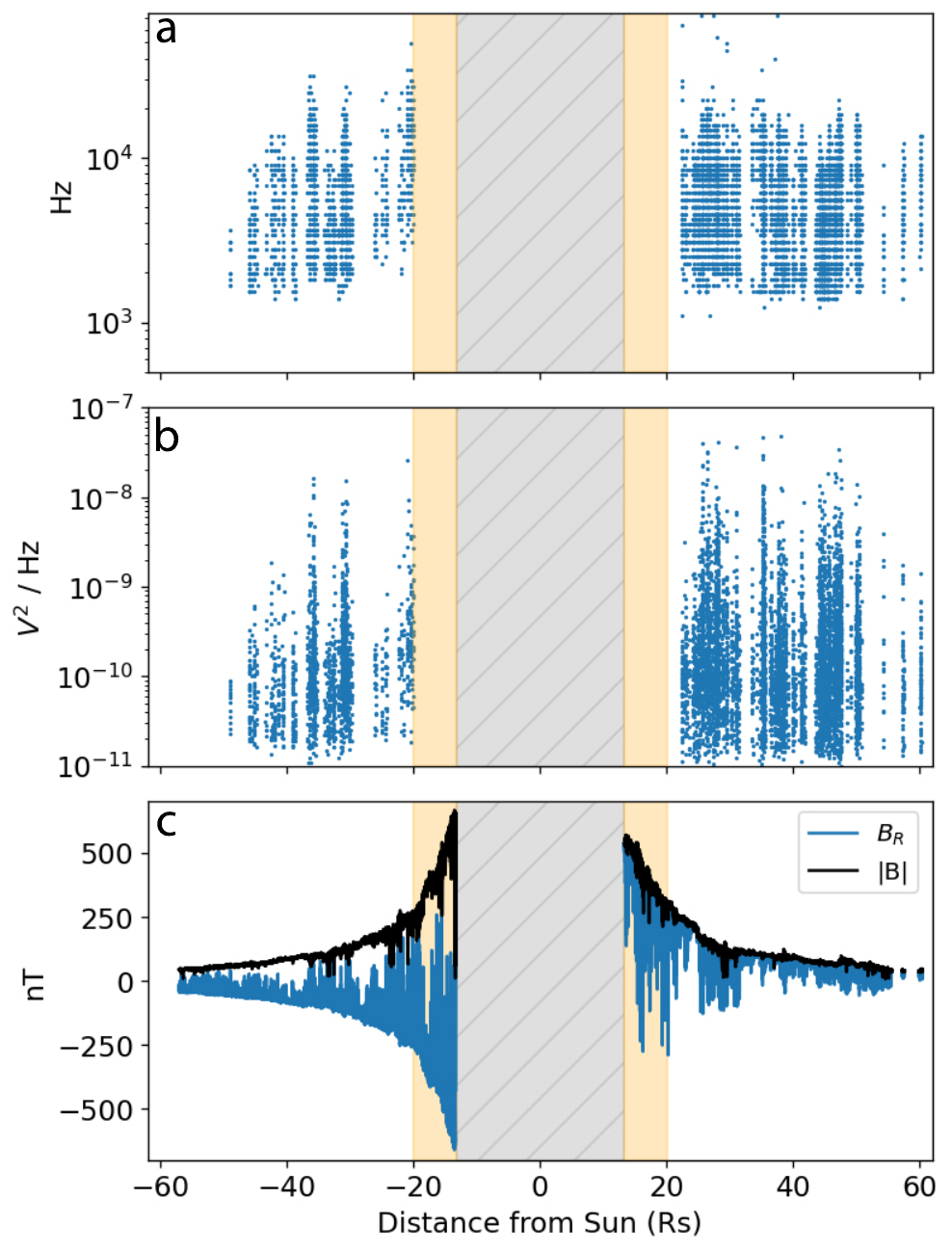}
\caption{ Distribution of wave properties for frequency-dispersed ion acoustic waves over Parker Solar Probe Encounter 11 as a function of distance from the Sun.  Negative (Positive) distances indicate inbound (outbound) portions of the orbit.  Radial distances that Parker Solar Probe did not visit on this solar encounter are shown by the gray hashed area.  Radial distances where it was not possible to distinguish frequency-dispersed ion acoustic waves from spacecraft wake-generated waves are indicated by the orange shaded region.  (a) Wave frequency (at highest wave amplitude). (b) Wave peak amplitude.  (c) Radial component of the background magnetic field (blue) and  background magnetic field magnitude (black). }
\label{fig_05}
\end{figure}

As an example, Figure \ref{fig_05} shows a distribution of wave properties over encounter 11, as a function of distance from the Sun.  The inbound (outbound) portion of the orbit is labeled with negative (positive) distance. Figures \ref{fig_05}a and \ref{fig_05}b show wave frequency and amplitude, respectively.  Figure \ref{fig_05}c shows the magnitude of the background magnetic field in black, and the radial component of the magnetic field in blue. This format allows heliospheric current sheet crossings \citep{Szabo2020} and switchback patches \citep{Bale2021} to be readily identified.  The gray hashed region indicates radial distances that Parker Solar Probe did not visit on this encounter, while the orange shaded region indicates regions where waves generated by the spacecraft plasma wake dominated the spectra such that frequency-dispersed ion acoustic waves could not be identified.  No clear radial trend is present in either the peak amplitude or peak frequency of the frequency-dispersed ion acoustic waves.  The spread in either quantity at any radial distance is much larger than their variability with radial distance. 


Table \ref{table_01} lists the number of frequency-dispersed ion acoustic waves detected on each encounter.  Earlier encounters have more wave identifications, but this may not be a physical effect.  There are several factors that complicate determination of a possible trend in radial occurrence:  (i) the orbital velocity of Parker Solar Probe and the speed of the solar wind vary strongly throughout any given orbit of Parker Solar Probe, causing the amount of solar wind plasma traversed during any 15 minute to be non-uniform, (ii) the data set includes many different Parker Solar Probe orbits, each of which has a different perihelion distance and orbital trajectory, (iii) the amount of time where wave detections can be made depends strongly on the occurrence of the plasma wake waves, which increases closer to the Sun.  

\begin{center}
\begin{tabular}{ c r |cr r }
 Encounter \# & \# Wave IDs & Encounter \# & \# Wave IDs \\ 
 \hline
 1 & N/A      & 9   & 17,143 \\  
 2 & 25,465 & 10 & 9,257   \\
 3 & 10,149 & 11 & 7,538   \\
 4 & 27,814 & 12 & 7,181   \\
 5 & 34,577 & 13 & 2,520   \\
 6 & 14,518 & 14 & 4,806  \\
 7 & 18,678 & 15 & 3,491 \\
 8 &   9,951 & 16 & 7,156 \\
\label{table_01}
\end{tabular}
\end{center}

\subsection{Derived Wave Properties} 
\label{sec:wave_statistics}

While time and frequency resolution limitations of the survey spectral data result in large variability for any given frequency-dispersed ion acoustic wave observation, the large number of wave detections makes it possible to extract wave properties statistically.  Several wave properties are determined this way, including the plasma-frame frequency of these waves, a distribution of wave numbers ($k$) and the wave-particle resonance energy most likely for wave growth. 

To determine the plasma frame wave frequency, the observed frequency ($f_{obs}$) of Doppler shifted waves is considered, 

\begin{equation}
f_{obs} = f_{pf} + \frac{\vec{k} \cdot \vec{v_{sw}}}{2\pi}
\end{equation}

for wave vector $\vec{k}$, solar wind velocity ($|\vec{v}_{sw}|$) in the frame of the spacecraft and plasma frame frequency $f_{pf}$. This equation can be normalized by the proton plasma frequency ($f_{pi} = (1/2\pi) \sqrt{ (q_p^2 n_p) / (m_p \epsilon_0)}$, where $q_p$ and $m_p$ are the charge and mass of a proton, the free space permittivity is $\epsilon_0$ and $n_p$ is plasma density). 

\begin{equation}
\frac{f_{obs}}{f_{pi}} = \frac{f_{pf}}{f_{pi}} + \frac{|k| |v_{sw}| cos(\theta_{k,vsw})}{2\pi f_{pi}}
\label{eqn_02a}
\end{equation}

For ion-acoustic waves, $\vec{k}$ is approximately along the background magnetic field ($\vec{B}$), therefore $cos(\theta_{k,vsw})$ is approximately $cos(\theta_{B,vsw})$.  Equation \ref{eqn_02a} can then be re-written,   

\begin{equation}
\frac{f_{obs}}{f_{pi}} = C + D cos(\theta_{B,vsw})
\label{eqn_0x}
\end{equation}

for $C =f_{pf} / f_{pi}$ and $D = (|k| |v_{sw}|) / (2\pi f_{pi})$. 

\begin{figure}[ht!]
\includegraphics[width = 89mm]{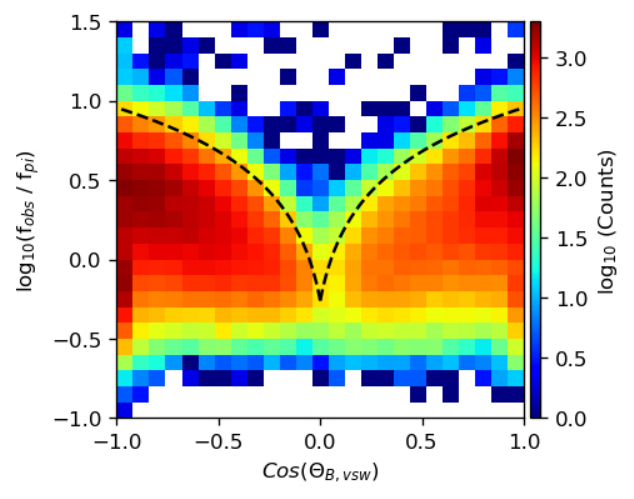}
\caption{ Number of frequency-dispersed ion acoustic wave observations as a function of wave frequency (normalized to the proton plasma frequency) and the cosine of the angle between the background magnetic field and the plasma flow vector in the frame of the spacecraft, $cos(\theta_{B,v_{sw}})$. The black dashed line describes a linear relationship between the normalized frequency and angle cosine (see text for details) that provides an upper bound to the bulk of the observations.}
\label{fig_07}
\end{figure}

Figure \ref{fig_07} shows the distribution of wave observations as a function of $f_{obs} / f_{pi}$ and $cos(\theta_{B,vsw})$.  A black dashed curve shows $f_{obs} / f_{pi} = C \pm D cos(\theta_{B,vsw})$, for $C = 0.55$ and $D = 8.5$.  There is a broad distribution of frequencies for each angle, but this is expected given that the wave signals disperse in frequency rapidly compared to a survey spectra averaging time.  However, the black curves provide an upper bound to the distribution.  As $cos(\theta_{B,vsw})$ approaches 0, the Doppler shift frequency contribution approaches 0 and the observed frequency approaches the plasma frame frequency.  Therefore, the data shown in Figure \ref{fig_07} indicate that the plasma frame frequency of the frequency-dispersed ion acoustic waves is $f_{pf} \approx 0.55 f_{pi}$.    


The Doppler shift equation can then be solved for $|k|$ as, 

\begin{equation}
|k| = \frac{ f_{obs} - f_{pf} }{ |v_{sw}| cos(\theta_{B,vsw}) } 
\label{eqn_02}
\end{equation}

\begin{figure}[ht!]
\includegraphics[width = 89mm]{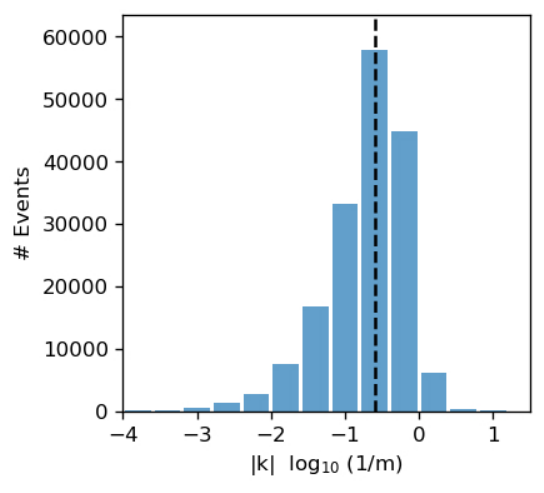}
\caption{ Number of frequency-dispersed ion acoustic wave observations as a function of wave number determined by Equation \ref{eqn_02}. }
\label{fig_08}
\end{figure}

Because $f_{obs}$, $|v_{sw}|$, and $cos(\theta_{B,vsw})$ are measured, this equation provides an estimate of $|k|$ if $f_{pf}$ is known.  Figure \ref{fig_08} shows the distribution of $|k|$ determined using this equation and the estimated plasma frame wave frequency determined from Figure \ref{fig_07}. This distribution is strongly peaked near $0.25$ $m^{-1}$, equivalent to a wavelength of $\sim 25$ m.  

As a consistency check, we use the definition of $D$ from Equation \ref{eqn_0x} together with the measured plasma parameters from Figure \ref{fig_01} to estimate $|k|$ for that event ($n_e = 612 \ cm^{-3}$, $|v_{sw}| = 305.7 \ km/s$).  Solving the definition of $D$ for $|k|$, using $D = 8.5$ from Figure \ref{fig_07}, $|k| = 0.905$, consistent with the second highest peak in Figure \ref{fig_08}.  


\begin{figure}[ht!]
\includegraphics[width = 89mm]{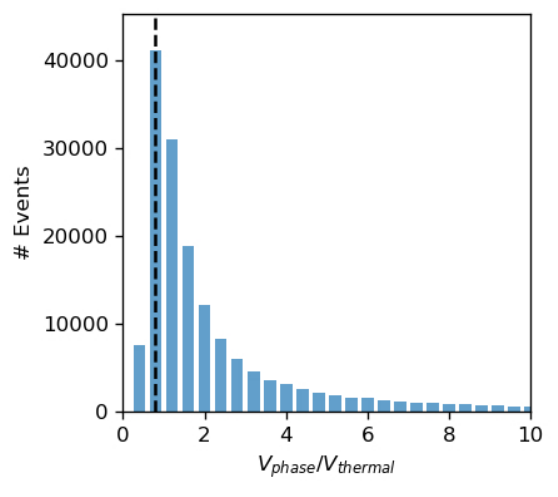}
\caption{ Number of frequency-dispersed ion acoustic wave observations as a function of inferred proton resonance speed normalized to the proton thermal speed. }
\label{fig_09}
\end{figure}

For an ion-acoustic wave driven by a Landau resonance, the speed of resonant ions parallel to $k$ will match the phase velocity of the fastest-growing wave frequency as $v_{res} = 2 \pi f_{pf} / |k|$. Using the distribution of $|k|$ from Figure \ref{fig_08} and $f_{pf} \approx 0.55 f_{pi}$, a distribution of $v_{res}$ is determined.  This distribution is shown in Figure \ref{fig_09}, where $v_{res}$ is normalized by the proton thermal speed $v_{th,p} = \sqrt{2 k_B T_p / m_p}$, where $k_B$ is the Boltzman constant and $T_p$ is the proton temperature.   This distribution is strongly peaked near 0.8 $v_{th,p}$.  

For a beam-plasma instability to drive Landau growth of these waves in this region of velocity space, the beams that drive these waves must have distinct properties.  (1) The beams must be slow.  That is, the most probable particle velocity in the beam should be approximately 1 proton thermal speed away from the peak of the thermal distribution.  (2) The beams must be cold, relative to the proton thermal speed, so that the distribution has a sharp enough positive slope to enable growth. (3) The beam instability must be short lived, to produce waves that appear in as $<< 1$s wave packets.    The next section examines proton distribution function observations for comparison with these expected beam properties. 


\section{Ion Signatures} 
\label{sec:ion_signatures}

The observed frequency-dispersed ion acoustic waves occur on timescales ($\sim$40 ms in Figure \ref{fig_01}), far faster than the measurement cadence of SPANi or SPC survey data ($\sim3$ to $\sim$30 s).  Therefore, it is unlikely to directly observe the unstable proton distributions driving these waves with these data.  However, if a proton beam is present and relaxes via Landau-resonant wave growth, the beam protons should spread in velocity space, creating a plateau-like structure in the distribution function.  Observed proton distribution functions can be examined for these remnant beam structures. 

\begin{figure}[ht!]
\includegraphics[width = 178mm]{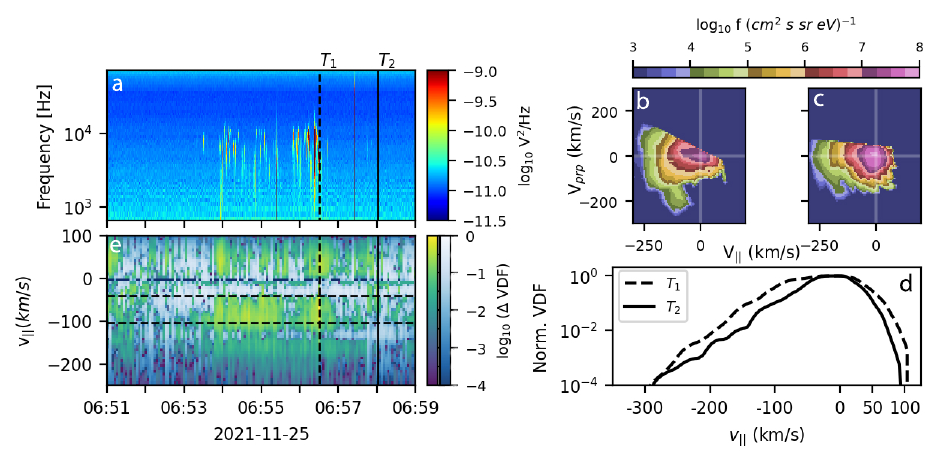}
\caption{ Comparison of proton distribution functions from intervals with and without  frequency-dispersed ion acoustic waves.  (a) Survey spectral data from the two differential electric field antenna pairs showing an interval of wave activity bounded by two intervals with weak or no wave activity.  (b) Proton distribution function from time $T_1$, as a function of velocity parallel and perpendicular to the background magnetic field direction, after the processing described in the text has been applied. (c) Same as (b), but from time $T_2$.  (d) Comparison of parallel slices through the proton distribution functions in (b) and (c), after the slices have been amplitude-normalized to their maximum value.  (e) The logarithm of the difference between amplitude-normalized parallel slices of the proton distribution functions at each time and the median of those slices from an interval when no waves are present, as a function of proton velocity and time (see text for details).  Positive difference values (more protons than the median slice) are shown with the yellow - purple colorbar. Negative values (fewer protons than the median slice) are shown with the white - blue colorbar. Horizontal dashed lines indicate 0.8 and 2 times the proton thermal speed.}
\label{fig_10}
\end{figure}

There are intervals of data where large numbers of frequency-dispersed ion acoustic waves appear in isolated patches.  These intervals are useful to explore plasma wave data and proton distribution functions from adjacent time periods with and without frequency-dispersed ion acoustic waves.  Figure \ref{fig_10} examines one such interval. 

Figure \ref{fig_10}a shows electric field power spectra ($V_{12}$ + $V_{34}$) for a $\sim8$ min interval on November 25, 2021.  At this time Parker Solar Probe was $\sim44.5$ $R_S$ from the Sun.  An interval of frequency-dispersed ion acoustic waves is evident, with intervals of no wave activity immediately before and after.  Figure \ref{fig_10}b and \ref{fig_10}c show proton distribution functions recorded by SPANi at the times $T_1$ and $T_2$, respectively (times indicated on Figure \ref{fig_10}a).  $T_1$ occurs during the wave interval, $T_2$ occurs outside the wave interval.


The proton distribution function data are processed in several steps.  The first steps follow \citet{Verniero2022}.  First, the measured energies and angles are converted to velocities, assuming mass and charge of a proton.  Next, the data are rotated from energy-angle space into a spherical coordinate system with angles $\theta$ (elevation) and $\phi$ (azimuth) in the SPANi instrument frame.  The resulting data are rotated into a magnetic-field aligned coordinate system, where the background magnetic field direction is defined using FIELDS fluxgate magnetometer data in the SPANi instrument frame, where the magnetometer data are low pass filtered and down sampled to the cadence of the SPANi measurements.  The basis vectors used for this coordinate system are $\hat{B}$, the magnetic field direction, $\hat{P} = \hat{B} \times \hat{N}$, and $\hat{Q} = \hat{B} \times \hat{P} $, where $\hat{N}$ is the SPANi instrument $+\hat{y}$ direction \citep{Livi2022}. 

The following additional data processing steps are then applied.  First, the distribution function peak is centered to (0,0,0) by subtracting off the three-dimensional solar wind velocity in the spacecraft frame.  This velocity is determined by treating the measured data points within the three-dimensional velocity space as a cloud of point masses, each with a mass equal to the proton flux at that point.  The center of mass for this point cloud is calculated and taken to be the solar wind velocity in the spacecraft frame.  Then, the principle axes of inertia for this cloud are determined, and the cloud is rotated into a frame where the eigenvector corresponding to the smallest eigenvalue (longest axis of the point cloud) is along $v_{||}$.  Here, $v_{||}$ and $v_{\perp}$ are velocities parallel and perpendicular to the background magnetic field.  For most distributions during this interval, the additional eigenvector rotation is small ($< 5^o$), because the rotation to magnetic-field aligned coordinates has already shifted the proton strahl \citep{Verniero2022} along $v_{||}$.  The additional eigenvector rotation carried out here has a more significant effect when the magnetic field is varying on timescales faster than the SPANi sample rate.  This effect has been examined in prior studies in the context of a `temperature broadening effect' \citep{Verscharen2011}.  At these times, the additional rotation can result in rotations of up to $\sim20^o$, producing improved alignment between the proton strahl and $v_{||}$.  The resulting point cloud is linearly interpolated onto a regular three-dimensional grid (128x128x128 pixels, linearly spaced over -400 km/s to +400 km/s in each dimension).  The data are then summed over $\pm 50$ km/s in each of the $v_{\perp}$ directions.  The result is a velocity distribution function slice along $v_{||}$ at each time.    

Figure \ref{fig_10}b and Figure \ref{fig_10}c show proton distribution functions from times $T_1$ and $T_2$, respectively.  Each proton distribution is plotted as a function of $v_{||}$ and $v_{\perp}$, after magnetic field alignment, centering, eigenvector rotation, interpolation onto the 3D grid, and summation over $\pm 50$ km/s in one of the $v_{\perp}$ directions.  The sharp diagonal boundaries through each distribution are caused by SPANi field of view limitations related to the Parker Solar Probe heat shield.  Figure \ref{fig_10}d shows $v_{||}$ slices for the distributions at  $T_1$ and $T_2$, where the distributions have both been normalized to their maximum value.  This normalization allows the relative shapes of the distribution functions to be compared.  The distribution at $T_1$ shows a broader shoulder of protons along $v_{||}$ than the distribution at $T_2$.  

Finally, Figure \ref{fig_10}e shows the time evolution of the $v_{||}$ slice shape through this interval.  Here, each $v_{||}$ slice is normalized to its maximum value.  Then, a `no waves' $v_{||}$ slice is defined as the median of the $v_{||}$ slice data at each speed over the first 45 s of  $v_{||}$ slices in this interval, when no waves are observed.  The difference is then calculated between each $v_{||}$ slice in the interval and the `no waves' $v_{||}$ slice.  This difference may be positive (more protons compared to the `no waves' slice) or negative (fewer protons compared to the `no waves' slice).  The base 10 logarithm of these differences is then plotted.  Positive values are plotted with the yellow-purple color bar while negative differences are plotted with the white-blue color bar.  During the plotted interval, there is an excess of protons between 0.8 $v_{th,p}$ and 2 $v_{th,p}$ (horizontal dashed lines) corresponding to times when the frequency-dispersed ion acoustic waves occur. When no waves are present, the proton excess at these velocities is reduced. This feature is interpreted as evidence of remnant proton beams; beams that have relaxed via the growth of plasma waves, leaving a `shoulder' on the proton distribution.  Further, the location in velocity space of the remnant beams overlaps with the most likely wave-driving beam velocities determined from wave properties, as show in Figure \ref{fig_09}.  

\section{Discussion} \label{sec:discussion}

Based on the presented analysis of plasma wave and proton velocity distribution function data, the most likely interpretation is that the observed frequency-dispersed ion acoustic waves are driven by impulsive cold proton beams.  These proton beams must be cold relative to the thermal proton distribution in order to produce wave growth near the estimated resonant speeds, $\sim v_{th,p}$ (Figure \ref{fig_09}, Figure \ref{fig_10}).  The beams must be impulsive because individual wave packets appear sporadically in time (Figure \ref{fig_02}, Figure \ref{fig_03}). 

The waves themselves have properties consistent with ion acoustic waves: electrostatic, wave vector along the background magnetic field, and frequencies close to $f_{pi}$. But, their plasma-frame frequency is below $f_{pi}$ ($\sim0.55 f_{pi}$), suggesting that they are more likely to be a type of beam-mode ion acoustic wave.  The wave frequency evolves much more rapidly ($>$ 500 kHz / s) than can be explained by solar wind velocity variability and resulting Doppler shift.  Two possible origins for the frequency variability are (i) evolution of the resonant velocity of the beam instability during wave growth / beam relaxation, or (ii) spatial / temporal variability of the proton beam energy due to a combination of spacecraft / solar wind motion and beam orientation in space.  The analysis presented here suggests that (i) is the smaller of the two effects.  The proton enhancement in Figure \ref{fig_10} spans at most a factor of 2 in proton velocity, suggesting that up to a factor of 2 in wave frequency shift can be explained by beam resonant velocity evolution.  The wave in Figure \ref{fig_01} shows a change in wave frequency by more than a factor of 3, while other waves in Figure \ref{fig_02} show frequency shifts of more than a factor of 10.   Further, some waves rise in frequency while some waves fall in frequency (Figure \ref{fig_02}).  Frequency shift due to a relaxing beam resonant velocity change should only allow decreasing frequency with time.  Therefore, effect (ii) is most likely the dominant effect causing the wave frequency shift. 

The origin of the proton beams responsible for the wave growth remains undetermined, but is of considerable interest.  What physical mechanisms can generate cold, impulsive, rapidly variable proton beams in the near-Sun solar wind far from shock fronts?  Possible mechanisms include (i) magnetic reconnection \citep{Dai2021}, (ii) steepened Alfv\'{e}n waves, or (iii) acceleration via Debye-scale electric potential structures.  

While it is beyond the scope of the current work to evaluate each mechanism in detail, (i) may be least likely given the scarcity of magnetic reconnection in the near-Sun solar wind (e.g. \citet{Fargette2023}) compared to number of wave observations (e.g. Figure \ref{fig_05}).  However, reconnection events may produce proton beams \citep{Phan2022} that travel far enough from the reconnection site that Parker Solar Probe would not detect the reconnection event itself.  Ideas $\it{ii}$ and $\it{iii}$ require detailed comparison between frequency-dispersed ion acoustic wave occurrences and occurrence of other plasma waves to prove or disprove.  This should be explored in future work.  With any of these mechanisms, proton beams are driven in the solar wind, frequency-dispersed ion acoustic waves grow in response, and the resulting proton distribution function is broadened in the parallel direction compared to when no waves are present.  We speculate that this broadening could be related to the evolution of the proton beam / shoulder feature in the near-Sun solar wind \citep{Marsch1982, Verniero2022}.   Also, the current study considered only proton velocity distribution functions.  The role of alpha particles remains to be explored \citep{McManus2024}.  

Finally, the survey plasma wave data used here have several limitations:  (i) they are averaged over times much longer than a given wave burst, (ii) they are pseudo-log averaged in frequency, and (iii) they are duty cycled such that they do not have full time coverage.  In spite of these limitations, they capture tens of thousands of wave events, and the aggregate statistics reveal a self-consistent picture.  Wave frequencies and polarization are consistent with beam-mode like ion acoustic waves, estimated resonance frequencies are consistent with proton distribution function measurements of remnant proton beam structures, and the waves show no clear trend in wave frequency and amplitude as one approaches the Sun. Remnant proton beam features were demonstrated to coincide with frequency-dispersed ion acoustic waves for one interval; however it is understood that this analysis may be impacted by the limited proton distribution function field of view coverage intrinsic to Parker Solar Probe.  Future work should examine the relationship between these waves and remnant beam structures over many such intervals.

\section{Conclusions} \label{sec:conclusions}

The data presented here demonstrate that Parker Solar Probe measures large numbers of frequency-dispersed ion acoustic waves throughout the near-Sun solar wind, on every solar encounter.  The waves were found to rapidly change frequency by a factor of 3x to 10x compared to their initial frequency, either from high to low or from low to high.  The data indicate that these waves are resonant with protons near the proton thermal speed.  It is suggested that these waves are most likely beam-mode ion acoustic waves, driven by cold, impulsive proton beams.  The origins of the proton beams require further investigation, and raise an interesting questions as to the physical mechanism responsible for impulsively accelerating protons throughout the near-Sun solar wind, and how that mechanism may tie to the global acceleration of the solar wind.    

\section{acknowledgments}
Parker Solar Probe was designed, built, and is now operated by the Johns Hopkins Applied Physics Laboratory as part of NASA's Living with a Star (LWS) program (contract NNN06AA01C). Support from the LWS management and technical team has played a critical role in the success of the Parker Solar Probe mission.  This work was also supported by NASA grant 80NSSC19K0305.  All data used here are publicly available on the FIELDS data archive: http://fields.ssl.berkeley.edu/data/ and the SWEAP data archive: http://sweap.cfa.harvard.edu/Data.html.


\begin{thebibliography}{}
\expandafter\ifx\csname natexlab\endcsname\relax\def\natexlab#1{#1}\fi
\providecommand{\url}[1]{\href{#1}{#1}}
\providecommand{\dodoi}[1]{doi:~\href{http://doi.org/#1}{\nolinkurl{#1}}}
\providecommand{\doeprint}[1]{\href{http://ascl.net/#1}{\nolinkurl{http://ascl.net/#1}}}
\providecommand{\doarXiv}[1]{\href{https://arxiv.org/abs/#1}{\nolinkurl{https://arxiv.org/abs/#1}}}

\bibitem[{{Anderson} {et~al.}(1981){Anderson}, {Eastman}, {Gurnett}, {Frank},
  \& {Parks}}]{Anderson1981}
{Anderson}, R.~R., {Eastman}, T.~E., {Gurnett}, D.~A., {Frank}, L.~A., \&
  {Parks}, G.~K. 1981, Journal of Geophysical Research, 86, 4493,
  \dodoi{10.1029/JA086iA06p04493}

\bibitem[{{Bale} {et~al.}(2016){Bale}, {Goetz}, {Harvey}, {Turin}, {Bonnell},
  {Dudok de Wit}, {Ergun}, {MacDowall}, {Pulupa}, {Andre}, {Bolton},
  {Bougeret}, {Bowen}, {Burgess}, {Cattell}, {Chandran}, {Chaston}, {Chen},
  {Choi}, {Connerney}, {Cranmer}, {Diaz-Aguado}, {Donakowski}, {Drake},
  {Farrell}, {Fergeau}, {Fermin}, {Fischer}, {Fox}, {Glaser}, {Goldstein},
  {Gordon}, {Hanson}, {Harris}, {Hayes}, {Hinze}, {Hollweg}, {Horbury},
  {Howard}, {Hoxie}, {Jannet}, {Karlsson}, {Kasper}, {Kellogg}, {Kien},
  {Klimchuk}, {Krasnoselskikh}, {Krucker}, {Lynch}, {Maksimovic}, {Malaspina},
  {Marker}, {Martin}, {Martinez-Oliveros}, {McCauley}, {McComas}, {McDonald},
  {Meyer-Vernet}, {Moncuquet}, {Monson}, {Mozer}, {Murphy}, {Odom},
  {Oliverson}, {Olson}, {Parker}, {Pankow}, {Phan}, {Quataert}, {Quinn},
  {Ruplin}, {Salem}, {Seitz}, {Sheppard}, {Siy}, {Stevens}, {Summers}, {Szabo},
  {Timofeeva}, {Vaivads}, {Velli}, {Yehle}, {Werthimer}, \&
  {Wygant}}]{Bale2016}
{Bale}, S.~D., {Goetz}, K., {Harvey}, P.~R., {et~al.} 2016, Space Science
  Reviews, 204, 49, \dodoi{10.1007/s11214-016-0244-5}

\bibitem[{{Bale} {et~al.}(2021){Bale}, {Horbury}, {Velli}, {Desai}, {Halekas},
  {McManus}, {Panasenco}, {Badman}, {Bowen}, {Chandran}, {Drake}, {Kasper},
  {Laker}, {Mallet}, {Matteini}, {Phan}, {Raouafi}, {Squire}, {Woodham}, \&
  {Woolley}}]{Bale2021}
{Bale}, S.~D., {Horbury}, T.~S., {Velli}, M., {et~al.} 2021, The Astrophysical
  Journal, 923, 174, \dodoi{10.3847/1538-4357/ac2d8c}

\bibitem[{{Balikhin} {et~al.}(2005){Balikhin}, {Walker}, {Treumann}, {Alleyne},
  {Krasnoselskikh}, {Gedalin}, {Andre}, {Dunlop}, \&
  {Fazakerley}}]{Balikhin2005}
{Balikhin}, M., {Walker}, S., {Treumann}, R., {et~al.} 2005, Geophysical
  Research Letters, 32, L24106, \dodoi{10.1029/2005GL024660}

\bibitem[{{Ber{\v{c}}i{\v{c}}} {et~al.}(2021){Ber{\v{c}}i{\v{c}}},
  {Maksimovi{\'c}}, {Halekas}, {Landi}, {Owen}, {Verscharen}, {Larson},
  {Whittlesey}, {Badman}, {Bale}, {Case}, {Goetz}, {Harvey}, {Kasper},
  {Korreck}, {Livi}, {MacDowall}, {Malaspina}, {Pulupa}, \&
  {Stevens}}]{Bercic2021}
{Ber{\v{c}}i{\v{c}}}, L., {Maksimovi{\'c}}, M., {Halekas}, J.~S., {et~al.}
  2021, The Astrophysical Journal, 921, 83, \dodoi{10.3847/1538-4357/ac1f1c}

\bibitem[{{Boardsen} {et~al.}(2015){Boardsen}, {Jian}, {Raines}, {Gershman},
  {Zurbuchen}, {Roberts}, \& {Korth}}]{Boardsen2015}
{Boardsen}, S.~A., {Jian}, L.~K., {Raines}, J.~L., {et~al.} 2015, Journal of
  Geophysical Research (Space Physics), 120, 10,207,
  \dodoi{10.1002/2015JA021506}

\bibitem[{{Bowen} {et~al.}(2020){Bowen}, {Mallet}, {Huang}, {Klein},
  {Malaspina}, {Stevens}, {Bale}, {Bonnell}, {Case}, {Chandran}, {Chaston},
  {Chen}, {Dudok de Wit}, {Goetz}, {Harvey}, {Howes}, {Kasper}, {Korreck},
  {Larson}, {Livi}, {MacDowall}, {McManus}, {Pulupa}, {Verniero}, \&
  {Whittlesey}}]{Bowen2020}
{Bowen}, T.~A., {Mallet}, A., {Huang}, J., {et~al.} 2020, The Astrophysical
  Journal Supplement, 246, 66, \dodoi{10.3847/1538-4365/ab6c65}

\bibitem[{{Carbone} {et~al.}(2021){Carbone}, {Sorriso-Valvo}, {Khotyaintsev},
  {Steinvall}, {Vecchio}, {Telloni}, {Yordanova}, {Graham}, {Edberg},
  {Eriksson}, {Johansson}, {V{\'a}sconez}, {Maksimovic}, {Bruno}, {D'Amicis},
  {Bale}, {Chust}, {Krasnoselskikh}, {Kretzschmar}, {Lorf{\`e}vre},
  {Plettemeier}, {Sou{\v{c}}ek}, {Steller}, {{\v{S}}tver{\'a}k},
  {Tr{\'a}vn{\'\i}{\v{c}}ek}, {Vaivads}, {Horbury}, {O'Brien}, {Angelini}, \&
  {Evans}}]{Carbone2021}
{Carbone}, F., {Sorriso-Valvo}, L., {Khotyaintsev}, Y.~V., {et~al.} 2021,
  Astronomy and Astrophysics, 656, A16, \dodoi{10.1051/0004-6361/202140931}

\bibitem[{{Case} {et~al.}(2020){Case}, {Kasper}, {Stevens}, {Korreck},
  {Paulson}, {Daigneau}, {Caldwell}, {Freeman}, {Henry}, {Klingensmith},
  {Bookbinder}, {Robinson}, {Berg}, {Tiu}, {Wright}, {Reinhart}, {Curtis},
  {Ludlam}, {Larson}, {Whittlesey}, {Livi}, {Klein}, \&
  {Martinovi{\'c}}}]{Case2020}
{Case}, A.~W., {Kasper}, J.~C., {Stevens}, M.~L., {et~al.} 2020, The
  Astrophysical Journal Supplement, 246, 43, \dodoi{10.3847/1538-4365/ab5a7b}

\bibitem[{{Cohen} {et~al.}(2020){Cohen}, {Cattell}, {Breneman}, {Davis},
  {Grul}, {Kersten}, {Wilson}, \& {Wygant}}]{Cohen2020}
{Cohen}, Z.~A., {Cattell}, C.~A., {Breneman}, A.~W., {et~al.} 2020, The
  Astrophysical Journal, 904, 174, \dodoi{10.3847/1538-4357/abbeec}

\bibitem[{{Dai} {et~al.}(2021){Dai}, {Wang}, \& {Lavraud}}]{Dai2021}
{Dai}, L., {Wang}, C., \& {Lavraud}, B. 2021, The Astrophysical Journal, 919,
  15, \dodoi{10.3847/1538-4357/ac0fde}

\bibitem[{{Davis} {et~al.}(2021){Davis}, {Cattell}, {Wilson}, {Cohen},
  {Breneman}, \& {Hanson}}]{Davis2021}
{Davis}, L.~A., {Cattell}, C.~A., {Wilson}, L.~B., I., {et~al.} 2021, The
  Astrophysical Journal, 913, 144, \dodoi{10.3847/1538-4357/abf56a}

\bibitem[{{Eastwood} {et~al.}(2005){Eastwood}, {Lucek}, {Mazelle}, {Meziane},
  {Narita}, {Pickett}, \& {Treumann}}]{Eastwood2005}
{Eastwood}, J.~P., {Lucek}, E.~A., {Mazelle}, C., {et~al.} 2005, Space Science
  Reviews, 118, 41, \dodoi{10.1007/s11214-005-3824-3}

\bibitem[{{Fargette} {et~al.}(2023){Fargette}, {Lavraud}, {Rouillard},
  {Houdayer}, {Phan}, {{\O}ieroset}, {Eastwood}, {Nicolaou}, {Fedorov},
  {Louarn}, {Owen}, \& {Horbury}}]{Fargette2023}
{Fargette}, N., {Lavraud}, B., {Rouillard}, A.~P., {et~al.} 2023, Astronomy and
  Astrophysics, 674, A98, \dodoi{10.1051/0004-6361/202346043}

\bibitem[{{Filbert} \& {Kellogg}(1979)}]{Filbert1979}
{Filbert}, P.~C., \& {Kellogg}, P.~J. 1979, Journal of Geophysical Reserach,
  84, 1369, \dodoi{10.1029/JA084iA04p01369}

\bibitem[{{Fox} {et~al.}(2016){Fox}, {Velli}, {Bale}, {Decker}, {Driesman},
  {Howard}, {Kasper}, {Kinnison}, {Kusterer}, \& {Lario}}]{Fox2016}
{Fox}, N.~J., {Velli}, M.~C., {Bale}, S.~D., {et~al.} 2016, Space Science
  Reviews, 204, 7, \dodoi{10.1007/s11214-015-0211-6}

\bibitem[{{Fuselier} \& {Gurnett}(1984)}]{Fuselier1984}
{Fuselier}, S.~A., \& {Gurnett}, D.~A. 1984, Journal of Geophysical Research,
  89, 91, \dodoi{10.1029/JA089iA01p00091}

\bibitem[{{Goodrich} {et~al.}(2019){Goodrich}, {Ergun}, {Schwartz}, {Wilson},
  {Johlander}, {Newman}, {Wilder}, {Holmes}, {Burch}, {Torbert},
  {Khotyaintsev}, {Lindqvist}, {Strangeway}, {Gershman}, \&
  {Giles}}]{Goodrich2019}
{Goodrich}, K.~A., {Ergun}, R., {Schwartz}, S.~J., {et~al.} 2019, Journal of
  Geophysical Research (Space Physics), 124, 1855, \dodoi{10.1029/2018JA026436}

\bibitem[{{Hellinger} {et~al.}(2006){Hellinger}, {Tr{\'a}vn{\'{\i}}{\v c}ek},
  {Kasper}, \& {Lazarus}}]{Hellinger2006}
{Hellinger}, P., {Tr{\'a}vn{\'{\i}}{\v c}ek}, P., {Kasper}, J.~C., \&
  {Lazarus}, A.~J. 2006, Geophysical Review Letters, 33, L09101,
  \dodoi{10.1029/2006GL025925}

\bibitem[{{Hull} {et~al.}(2006){Hull}, {Larson}, {Wilber}, {Scudder}, {Mozer},
  {Russell}, \& {Bale}}]{Hull2006}
{Hull}, A.~J., {Larson}, D.~E., {Wilber}, M., {et~al.} 2006, Geophysical
  Research Letters, 33, L15104, \dodoi{10.1029/2005GL025564}

\bibitem[{{Jian} {et~al.}(2009){Jian}, {Russell}, {Luhmann}, {Strangeway},
  {Leisner}, \& {Galvin}}]{Jian2009}
{Jian}, L.~K., {Russell}, C.~T., {Luhmann}, J.~G., {et~al.} 2009, The
  Astrophysical Journal Letters, 701, L105,
  \dodoi{10.1088/0004-637X/701/2/L105}

\bibitem[{{Jian} {et~al.}(2014){Jian}, {Wei}, {Russell}, {Luhmann}, {Klecker},
  {Omidi}, {Isenberg}, {Goldstein}, {Figueroa-Vi{\~n}as}, \&
  {Blanco-Cano}}]{Jian2014}
{Jian}, L.~K., {Wei}, H.~Y., {Russell}, C.~T., {et~al.} 2014, The Astrophysical
  Journal, 786, 123, \dodoi{10.1088/0004-637X/786/2/123}

\bibitem[{{Kasper} {et~al.}(2002){Kasper}, {Lazarus}, \& {Gary}}]{Kasper2002}
{Kasper}, J.~C., {Lazarus}, A.~J., \& {Gary}, S.~P. 2002, Geophysical Research
  Letters, 29, 1839, \dodoi{10.1029/2002GL015128}

\bibitem[{{Kasper} {et~al.}(2016){Kasper}, {Abiad}, {Austin}, {Balat-Pichelin},
  {Bale}, {Belcher}, {Berg}, {Bergner}, {Berthomier}, {Bookbinder}, {Brodu},
  {Caldwell}, {Case}, {Chand ran}, {Cheimets}, {Cirtain}, {Cranmer}, {Curtis},
  {Daigneau}, {Dalton}, {Dasgupta}, {DeTomaso}, {Diaz-Aguado}, {Djordjevic},
  {Donaskowski}, {Effinger}, {Florinski}, {Fox}, {Freeman}, {Gallagher},
  {Gary}, {Gauron}, {Gates}, {Goldstein}, {Golub}, {Gordon}, {Gurnee}, {Guth},
  {Halekas}, {Hatch}, {Heerikuisen}, {Ho}, {Hu}, {Johnson}, {Jordan},
  {Korreck}, {Larson}, {Lazarus}, {Li}, {Livi}, {Ludlam}, {Maksimovic},
  {McFadden}, {Marchant}, {Maruca}, {McComas}, {Messina}, {Mercer}, {Park},
  {Peddie}, {Pogorelov}, {Reinhart}, {Richardson}, {Robinson}, {Rosen},
  {Skoug}, {Slagle}, {Steinberg}, {Stevens}, {Szabo}, {Taylor}, {Tiu}, {Turin},
  {Velli}, {Webb}, {Whittlesey}, {Wright}, {Wu}, \& {Zank}}]{Kasper2016}
{Kasper}, J.~C., {Abiad}, R., {Austin}, G., {et~al.} 2016, Space Science
  Reviews, 204, 131, \dodoi{10.1007/s11214-015-0206-3}

\bibitem[{{Livi} {et~al.}(2022){Livi}, {Larson}, {Kasper}, {Abiad}, {Case},
  {Klein}, {Curtis}, {Dalton}, {Stevens}, {Korreck}, {Ho}, {Robinson}, {Tiu},
  {Whittlesey}, {Verniero}, {Halekas}, {McFadden}, {Marckwordt}, {Slagle},
  {Abatcha}, {Rahmati}, \& {McManus}}]{Livi2022}
{Livi}, R., {Larson}, D.~E., {Kasper}, J.~C., {et~al.} 2022, The Astrophysical
  Journal, 938, 138, \dodoi{10.3847/1538-4357/ac93f5}

\bibitem[{{Malaspina} {et~al.}(2016){Malaspina}, {Jaynes}, {Boul{\'e}},
  {Bortnik}, {Thaller}, {Ergun}, {Kletzing}, \& {Wygant}}]{Malaspina2016}
{Malaspina}, D.~M., {Jaynes}, A.~N., {Boul{\'e}}, C., {et~al.} 2016,
  Geophysical Review Letters, 43, 7878, \dodoi{10.1002/2016GL069982}

\bibitem[{{Malaspina} {et~al.}(2022{\natexlab{a}}){Malaspina}, {Tigik}, \&
  {Vaivads}}]{Malaspina2022_wakes}
{Malaspina}, D.~M., {Tigik}, S.~F., \& {Vaivads}, A. 2022{\natexlab{a}}, The
  Astrophysical Journal Letters, 936, L20, \dodoi{10.3847/2041-8213/ac8c8f}

\bibitem[{{Malaspina} {et~al.}(2022{\natexlab{b}}){Malaspina}, {Chasapis},
  {Tatum}, {Salem}, {Bale}, {Bonnell}, {Dudok de Wit}, {Goetz}, {Pulupa},
  {Halekas}, {Whittlesey}, {Livi}, {Case}, {Stevens}, \&
  {Larson}}]{Malaspina2022_KAW}
{Malaspina}, D.~M., {Chasapis}, A., {Tatum}, P., {et~al.} 2022{\natexlab{b}},
  The Astrophysical Journal, 936, 128, \dodoi{10.3847/1538-4357/ac87a7}

\bibitem[{{Malaspina} {et~al.}(2022{\natexlab{c}}){Malaspina}, {Stenborg},
  {Mehoke}, {Al-Ghazwi}, {Shen}, {Hsu}, {Iyer}, {Bale}, \& {Dudok de
  Wit}}]{Malaspina2022_dust}
{Malaspina}, D.~M., {Stenborg}, G., {Mehoke}, D., {et~al.} 2022{\natexlab{c}},
  The Astrophysical Journal, 925, 27, \dodoi{10.3847/1538-4357/ac3bbb}

\bibitem[{{Marsch} {et~al.}(1982){Marsch}, {Schwenn}, {Rosenbauer},
  {Muehlhaeuser}, {Pilipp}, \& {Neubauer}}]{Marsch1982}
{Marsch}, E., {Schwenn}, R., {Rosenbauer}, H., {et~al.} 1982, Journal of
  Geophysical Research, 87, 52, \dodoi{10.1029/JA087iA01p00052}

\bibitem[{{Matteini} {et~al.}(2007){Matteini}, {Landi}, {Hellinger},
  {Pantellini}, {Maksimovic}, {Velli}, {Goldstein}, \& {Marsch}}]{Matteini2007}
{Matteini}, L., {Landi}, S., {Hellinger}, P., {et~al.} 2007, Geophysical
  Research Letters, 34, L20105, \dodoi{10.1029/2007GL030920}

\bibitem[{{McManus} {et~al.}(2024){McManus}, {Klein}, {Bale}, {Bowen}, {Huang},
  {Larson}, {Livi}, {Rahmati}, {Romeo}, {Verniero}, \&
  {Whittlesey}}]{McManus2024}
{McManus}, M.~D., {Klein}, K.~G., {Bale}, S.~D., {et~al.} 2024, The
  Astrophysical Journal, 961, 142, \dodoi{10.3847/1538-4357/ad05ba}

\bibitem[{{Mozer} {et~al.}(2023){Mozer}, {Bale}, {Kellogg}, {Romeo}, {Vasko},
  \& {Verniero}}]{Mozer2023}
{Mozer}, F., {Bale}, S., {Kellogg}, P., {et~al.} 2023, Physics of Plasmas, 30,
  062111, \dodoi{10.1063/5.0151423}

\bibitem[{{Mozer} {et~al.}(2021){Mozer}, {Vasko}, \& {Verniero}}]{Mozer2021}
{Mozer}, F.~S., {Vasko}, I.~Y., \& {Verniero}, J.~L. 2021, The Astrophysical
  Journal Letters, 919, L2, \dodoi{10.3847/2041-8213/ac2259}

\bibitem[{{Phan} {et~al.}(2022){Phan}, {Verniero}, {Larson}, {Lavraud},
  {Drake}, {{\O}ieroset}, {Eastwood}, {Bale}, {Livi}, {Halekas}, {Whittlesey},
  {Rahmati}, {Stansby}, {Pulupa}, {MacDowall}, {Szabo}, {Koval}, {Desai},
  {Fuselier}, {Velli}, {Hesse}, {Pyakurel}, {Maheshwari}, {Kasper}, {Stevens},
  {Case}, \& {Raouafi}}]{Phan2022}
{Phan}, T.~D., {Verniero}, J.~L., {Larson}, D., {et~al.} 2022, Geophysical
  Research Letters, 49, e96986,
  \dodoi{10.1029/2021GL09698610.1002/essoar.10508706.1}

\bibitem[{{Szabo} {et~al.}(2020){Szabo}, {Larson}, {Whittlesey}, {Stevens},
  {Lavraud}, {Phan}, {Wallace}, {Jones-Mecholsky}, {Arge}, {Badman},
  {Odstrcil}, {Pogorelov}, {Kim}, {Riley}, {Henney}, {Bale}, {Bonnell}, {Case},
  {Dudok de Wit}, {Goetz}, {Harvey}, {Kasper}, {Korreck}, {Koval}, {Livi},
  {MacDowall}, {Malaspina}, \& {Pulupa}}]{Szabo2020}
{Szabo}, A., {Larson}, D., {Whittlesey}, P., {et~al.} 2020, The Astrophysical
  Journal Supplement, 246, 47, \dodoi{10.3847/1538-4365/ab5dac}

\bibitem[{{Tigik} {et~al.}(2022){Tigik}, {Vaivads}, {Malaspina}, \&
  {Bale}}]{Tigik2022}
{Tigik}, S.~F., {Vaivads}, A., {Malaspina}, D.~M., \& {Bale}, S.~D. 2022, The
  Astrophysical Journal (submitted), .

\bibitem[{{Vasko} {et~al.}(2022){Vasko}, {Mozer}, {Bale}, \&
  {Artemyev}}]{Vasko2022}
{Vasko}, I.~Y., {Mozer}, F.~S., {Bale}, S.~D., \& {Artemyev}, A.~V. 2022,
  Geophysical Research Letters, 49, e98640, \dodoi{10.1029/2022GL098640}

\bibitem[{{Vech} {et~al.}(2021){Vech}, {Malaspina}, {Cattell}, {Schwartz},
  {Ergun}, {Klein}, {Kromyda}, \& {Chasapis}}]{Vech2021}
{Vech}, D., {Malaspina}, D.~M., {Cattell}, C., {et~al.} 2021, Journal of
  Geophysical Research (Space Physics), 126, e29221,
  \dodoi{10.1029/2021JA029221}

\bibitem[{{Verniero} {et~al.}(2020){Verniero}, {Larson}, {Livi}, {Rahmati},
  {McManus}, {Pyakurel}, {Klein}, {Bowen}, {Bonnell}, {Alterman}, {Whittlesey},
  {Malaspina}, {Bale}, {Kasper}, {Case}, {Goetz}, {Harvey}, {Korreck},
  {MacDowall}, {Pulupa}, {Stevens}, \& {de Wit}}]{Verniero2020}
{Verniero}, J.~L., {Larson}, D.~E., {Livi}, R., {et~al.} 2020, The
  Astrophysical Journal Supplement, 248, 5, \dodoi{10.3847/1538-4365/ab86af}

\bibitem[{{Verniero} {et~al.}(2022){Verniero}, {Chandran}, {Larson}, {Paulson},
  {Alterman}, {Badman}, {Bale}, {Bonnell}, {Bowen}, {de Wit}, {Kasper},
  {Klein}, {Lichko}, {Livi}, {McManus}, {Rahmati}, {Verscharen}, {Walters}, \&
  {Whittlesey}}]{Verniero2022}
{Verniero}, J.~L., {Chandran}, B.~D.~G., {Larson}, D.~E., {et~al.} 2022, The
  Astrophysical Journal, 924, 112, \dodoi{10.3847/1538-4357/ac36d5}

\bibitem[{{Verscharen} {et~al.}(2019){Verscharen}, {Klein}, \&
  {Maruca}}]{Verscharen2019}
{Verscharen}, D., {Klein}, K.~G., \& {Maruca}, B.~A. 2019, Living Reviews in
  Solar Physics, 16, 5, \dodoi{10.1007/s41116-019-0021-0}

\bibitem[{{Verscharen} \& {Marsch}(2011)}]{Verscharen2011}
{Verscharen}, D., \& {Marsch}, E. 2011, Annales Geophysicae, 29, 909,
  \dodoi{10.5194/angeo-29-909-2011}

\bibitem[{{Wicks} {et~al.}(2016){Wicks}, {Alexander}, {Stevens}, {Wilson},
  {Moya}, {Vi{\~n}as}, {Jian}, {Roberts}, {O'Modhrain}, {Gilbert}, \&
  {Zurbuchen}}]{Wicks2016}
{Wicks}, R.~T., {Alexander}, R.~L., {Stevens}, M., {et~al.} 2016, The
  Astrophysical Journal, 819, 6, \dodoi{10.3847/0004-637X/819/1/6}

\bibitem[{{Wilson} {et~al.}(2014){Wilson}, {Sibeck}, {Breneman}, {Le Contel},
  {Cully}, {Turner}, {Angelopoulos}, \& {Malaspina}}]{Wilson2014}
{Wilson}, L.~B., {Sibeck}, D.~G., {Breneman}, A.~W., {et~al.} 2014, Journal of
  Geophysical Research (Space Physics), 119, 6475, \dodoi{10.1002/2014JA019930}

\end{thebibliography}


\end{document}